\documentclass[12pt]{article}
\usepackage{epsf}
\usepackage{amsmath}

\usepackage{graphics}
\usepackage{cite}

\begin{titlepage}

\begin{document}

\hfill February 2021\\

\begin{center}

{\bf \Large Ultralight Axions Versus Primordial Black Holes \\
}
\vspace{2.5cm}
{\bf Claudio Corian\`{o}\footnote{claudio.coriano@le.infn.it} and Paul H. Frampton\footnote{ paul.h.frampton@gmail.com} \\  }
\vspace{0.5cm}
{\it Dipartimento di Matematica e Fisica "Ennio De Giorgi",\\ 
Universit\`{a} del Salento and INFN-Lecce,\\ Via Arnesano, 73100 Lecce, Italy}\\
\vspace{0.5cm}
{\bf Jihn E. Kim\footnote{jihnekim@gmail.com}}\\
\vspace{0.5cm}
{\it Department of Physics, Kyung Hee University, 26 Gyungheedaero, \\
Dongdaemun-Gu, Seoul 02447, Republic of Korea}

\begin{abstract}
\noindent
We reconsider entropy arguments which have been previously argued
to support the idea that the dark matter constituents are primordial
black holes with many solar masses. It has recently been shown that
QCD axions which solve the strong CP problem may have masses
$m_a$ in the extended range $10^{-3}$eV $> m_a > 10^{-33}$ eV. Ultralight axions
provide so many degrees of freedom that their entropy can exceed that
of primordial black holes. This suggests that ultralight
axions are more suited than primordial black holes to be
constituents of dark matter.

\end{abstract}

\end{center}
\end{titlepage}

\section{Introduction} 

\noindent
One motivation for new physics beyond the standard model
of particle theory is
the strong CP problem. The absence of observed CP violation in strong
interactions dictates that a parameter $\bar{\Theta}$ defined in terms of
$\Theta$ occurring in the kinetic part of the QCD lagrangian
\begin{equation}
{\cal L}_{QCD}^{kinetic} = - \frac{1}{4} G_{\mu\nu}^a G^{\mu\nu,a} -
\frac{\Theta}{64\pi^2} \epsilon^{\mu\nu\alpha\beta} G_{\mu\nu}^a G_{\alpha\beta}^a
\label{QCDkinetic}
\end{equation}
\noindent
by combining with a mass matrix phase 
\begin{equation}
\bar{\Theta} = \Theta +\arg \det \left( M_uM_d \right)
\label{ThetaBar}
\end{equation}
to be invariant under chiral transformations on the quarks,
satisfy
\begin{equation}
\bar{\Theta} < 10^{-10}
\label{tuning}
\end{equation}

\bigskip

\noindent
This is the strong CP problem and the choice is {\it either} to accept that the
fine-tuning in Eq.(\ref{tuning}) as no more bothersome than {\it e.g.} 
\begin{equation}
\left| \frac{M_{\nu}}{M_{top}} \right| < 10^{-12}
\label{fermiontuning}
\end{equation}
{\it or} to adopt the orthodoxy which predicts axions \cite{Weinberg,Wilczek,Kim1,SVZ,Zhitnitsky,DFS } recently generalised to an ultralight axion (ULA) \cite{Hook1,Hook2,Ringwald1,Ringwald2,Kim2} with mass in an extended range
\begin{equation}
10^{-33} \textrm{eV} \leq m_a \leq 10^{-3} \textrm{eV}.
\label{axionmass}
\end{equation}
Interestingly, this includes the mass invoked in \cite{Kim3,HOTWitten} to ameliorate
problems with cold dark matter.

\bigskip

\noindent
In the present article, we shall study ULAs  as dark matter using thermodynamic arguments. 

\newpage

\section{PBHs and ULAs}

\noindent
Before discussing the nature of dark matter {\it i.e.} what are its constituents,
we shall discuss the origin of dark matter {\it i.e} the reason why it exists.
Since the first discoveries of dark matter in clusters \cite{Zwicky} then in 
individual galaxies \cite{Rubin}, a chaotic array of candidates, ranging in
mass by over 100 orders of magnitude, have been proposed. To bring order
from this chaos requires a guiding principle.

\bigskip

\noindent
We shall assume that this guiding principle arises from thermodynamics. The
early universe when the dark matter originated will be treated\cite{FramptonDM}  as a
thermodynamic system which can be dealt with using the method
which was first presented by Boltzmann\cite{Boltzmann}. To be a
suitable for such a treatment, there are three basic requirements:\\
(1) The system contains a very large number of particles.\\
(2) The system is close to thermal equilibrium.\\
(3) The system is thermodynamically isolated.\\

\noindent
Condition (1):  Boltzmann regarded Avagadro's number $N_A\sim 10^{24}$ 
as a very large number. The number of particles in the universe is $N_U \sim 10^{80}$,
far larger than $N_A$, so this condition is easily satisfied.\\
Condition (2) appears to be satisfied because the CMB spectrum is the best black-body
spectrum ever measured, strongly suggesting excellent thermal equilibrium
in the early universe.\\
Condition (3) is the most subtle. We admit that, although this seems a very reasonable
assumption, it is debatable. The reader could ask about the possibility that the universe 
interacts, or is entangled, with other universes. Could that not allow energy to flow in to, and out of, the
universe? Indeed it could, so we exclude such scenarios.

\bigskip

\noindent
Provided that the early universe can be treated as a thermodynamic system
close to thermal equilibrium, it follows both that the entropy of the universe is
a sensible concept and that because of Boltzmann's second law the entropy
will only remain constant or increase. This suggests that we choose as our 
\\

\noindent
{\bf Guiding principle}\\
{\it Subject to dynamical constraints, the dark matter constituents are such that
the entropy of the universe reaches its maximum possible value}.

\newpage

\noindent
A brief note about notation. To abbreviate the exponents  in very large
numbers it will be useful to employ the symbol ${\cal G} \equiv 10^{100}$,
For the ULAs our mass unit is the electron volt, $eV$. For PBHs our mass
unit is the Solar mass, $M_{\odot}$. They are related by $M_{\odot}\sim 10^{66}$ eV.
In our discussion, we shall sometimes ignore coefficients $O(1)$ and
equations will use $\sim$, not $=$. Conclusions 
derived therefrom about physics will, nevertheless, be ironclad.

\bigskip

\noindent
Let $S_U$ be the total dimensionless entropy, that is the Boltzmann entropy
divided by Boltzmann's constant, of the universe, excluding it surface. 

\bigskip

\noindent
The surface entropy equals its area in Planck units. For radius $R(t_0) \sim 44\, Gly$, the surface entropy is $\sim 10^{23}\,{\cal G}$.
According to the holographic principle of 't Hooft\cite{Hooft}, the highest possible entropy
of the universe, whatever its content, is the entropy of its boundary surface. This provides the maximum
$S_U(t_0) \leq 10^{23}\,{\cal G}$. This inequality is saturated, again according to
't Hooft\cite{Hooft}, only if the universe
itself is a black hole which is not possible because the radius $R(t_0) \sim 44\,Gly$ exceeds
its Schwarzschild radius $R_{Sch}(t_0) \sim 30\, Gly$ by roughly a factor $\sim \sqrt{2}$, to be used later. Thus $S_U(t_0) < 10^{23}\,{\cal G}$. A perspicuous discussion of entropy is in \cite{Lenny}.

\bigskip

\noindent
Let us consider a toy model for the visible universe, sufficiently realistic for the present 
purpose. In this model there are exactly $10^{11}$ galaxies, each with mass exactly $10^{12}M_{\odot}$.

\bigskip

\noindent
The present universe has, excluding the dark matter, a number of components including
the CMB, relic neutrinos, stars and black holes. According to {\it e.g.} Weinberg's
textbook\cite{Weinberg2}, the first three contribute to $S_U(t_0)$ by the amounts $\sim 10^{-12} \,{\cal G}$, $\sim 10^{-12}\, {\cal G}$,
$\sim 10^{-20}\, {\cal G}$ respectively. 

\bigskip

\noindent
It is believed that each galaxy contains near to its center a
supermassive black hole (SMBH). In our toy model, we shall assume that all $10^{11}$ of
these SMBHs have mass exactly $10^7 M_{\odot}$. The Bekenstein-Hawking entropy\cite{Bekenstein,Hawking}
of a black hole with mass $\eta M_{\odot}$ is $S_{BH}(\eta M_{\odot}) \sim 10^{78} \eta^2$. This
leads to a contribution to $S_U(t_0)$ of $\sim 10^{(78+14+11)} =10^3 \,{\cal G}$, over a trillion times more than
the contributions from CMB, relic neutrinos and stars, all added together. \\

\noindent
We assume the dark
energy does not contribute to $S_U$, as is the case if it is described by a cosmological constant 
with no degrees of freedom. All observations of dark energy are currently consistent with
this interpretation.

\bigskip

\noindent
The above discussion implies that without dark matter $S_U(t_0)$ 
should lie between the bounds
\begin{equation}
10^3\, {\cal G} \leq S_U(t_0) < 10^{23}\, {\cal G}
\label{bounds}
\end{equation}

\newpage

\noindent
Our guiding principle {\it ut supra} requires that, when dark matter is added, the
value of $S_U(t_0)$ will be as near to the upper limit in Eq.(\ref{bounds})
as possible. We believe this is what determines the choice by Nature for
what is the dark matter candidate. How close to the 't Hooft holographic
limit can the dark matter take us?

\bigskip

\noindent
To answer this question, which is central to the present Letter, we revisit
two facts about the visible universe that its radius $R(t_0) \sim 44\, Gly$ while
its Schwarzschild radius $R_{Sch}(t_0) \sim 30\, Gly$. We propose that
the dark matter provides enough entropy to saturate the 't Hooft bound
{\it as if the universe were a black hole}! We thus use $R(t_0) \sim
30\, Gly$ and the holographic bound is reduces by a number $\sim 2$
which we shall absorb within the order of magntude.  Therefore, we
require that, if possible, the dark matter provide a contribution $\sim 10^{23}\, {\cal G}$.

\bigskip

\noindent
In subsections (2.1) and (2.2), we shall study whether such a large contribution as
$\sim 10^{23} {\cal G}$ to $S_U(t_0)$
is possible for, respectively, primordial black holes and ultralight axions.

\subsection{Primordial black holes}

\noindent
It was hypothesised over five years ago \cite{PaulFrampton}
(see also \cite{GeorgeChapline,Chapline2,ChaplineBarbieri,FKTY,ChaplineFrampton})
that the best candidate for dark matter at that time might be intermediate mass black holes,
summarised by the acronym $DM=PIMBHs$. One motivation came from the lack of
any convincing particle theory candidate. A second motivation came from the fact that
$DM=PIMBHs$ requires no new physics beyond the standard model of particle
theory plus general relativity. Note that the $PIMBHs$  must be primordial because
there is insufficient baryonic matter to accommodate all of the dark matter. 
A third motivation came from entropy considerations, discussed {\it ut infra}.
At that time, axions which solve the strong CP problem of QCD were believed to have
mass $m_a \geq 10^{-12} eV$. A recent extension of this lower limit
on $m_a$ , see subsection 2.2 {\it ut infra}, was the inspiration for our reconsideration
of the entropy of the universe.

\bigskip

\noindent
The paper \cite{PaulFrampton} was published more than four months before
the announcemnent of the discovery by LIGO\cite{LIGO} of gravitational waves.
That dramatic discovery revealed that black holes exist with tens of $M_{\odot}$'s, 
although they are
probably not primordial, rather the result of gravitational collapse followed by
mergers.
Initially, we were interested only in dark matter within the Milky Way, since this was
the most accessible to detection by microlensing \cite{MACHO}.

\bigskip

\noindent
Within the Milky Way, stability of the galactic disk\cite{Ostriker} puts an upper limit
on the allowed mass of PIMBHs. Although the Milky Way has an SMBH at its centre,
$Sgr A^*$, with mass $M\sim 4\times 10^6 M_{\odot}$, it is impossible that
there exists a second such object in the spiral arms because, as shown in
\cite{Ostriker} by numerical simulation, this would have
destabilised the galactic disk within its age.
To stay away from such a disaster, in \cite{PaulFrampton} we discussed the
range of masses
\begin{equation}
10^2 M_{\odot} \leq M_{PIMBH} \leq 10^5 M_{\odot}
\label{PIMBHmass}
\end{equation}
where the upper limit was inspired by \cite{Ostriker} .

\bigskip

\noindent
The lower limit in Eq. (\ref{PIMBHmass}) corresponds to
that in Eq. (\ref{bounds}), {\it ut supra}, as follows.
Let us suppose the dark matter in our toy model of the universe is entirely
made up of black holes each with mass $10^p M_{\odot}$. Using the total
dark matter mass $10^{23} M_{\odot}$ and the Bekenstein-Hawking formula
then gives a contribution to $S_U(t_0)$ which is
\begin{equation}
S_U (t_0) \sim 10^{(23-p+78+2p)} = 10^{(1+p)}\, {\cal G}
\label{entropyPIMBHs}
\end{equation}
so that for $p=2$ this contribution
coincides with the lower bound for the entropy of the universe in Eq.(\ref{bounds}).
As we increase $p=2$ to $p=5$ as suggested in Eq.(\ref{PIMBHmass}) the
entropy of the universe increases by 3 orders of magnitude (OoMs).

\bigskip

\noindent
In \cite{PaulFrampton}, we were satisfied to increase $DM=PIMBHs$ 
by just 3 OoMs. But the 't Hooft holographic principle, Eq.(\ref{bounds}) suggests
that the increase could be far larger, by $\sim 20$ OoMs.
Is this possible by dark matter constituents which are PBHs?
The upper limit in Eq.(\ref{PIMBHmass}) applies only inside the Milky Way.
Outside, the disk stability constraint is inapplicable and heavier black holes are
permitted. Of course, we know about the SMBHs, one per galaxy, which
were used to obtain the lower bound in Eq.(\ref{bounds}). But
what if all the dark matter were made from additional black holes of
mass $\sim 10^p M_{\odot}$ with $p>5$?

\bigskip

\noindent
We know there is dark matter associated with clusters, and it is unknown
whether there are stupendously large black holes (SLABs) \cite{Carr}
unassociated with either galaxies or clusters. The highest $p$-value
considered in \cite{Carr} is $p=18$, a black hole with one hundred
thousandth of the total mass of the universe. However unlikely, and
contrary to the need for dark matter within galaxies, let us assume
that there is a monochromatic spectrum at $p=18$, 
in order to maximise the entropy of the universe. From 
Eq.(\ref{entropyPIMBHs}) we find $S_U(t_0)\sim 10^{19}\,{\cal G}$
which, in Eq.(\ref{bounds}), succeeds in acquiring only 16 of the
20 desired OoMs. Thus, saturation of the 't Hooft holographic upper
limit seems impossible, using only PBHs. 

\bigskip

\noindent
In the next subsection, by considering ultralight axions, 
we shall find a significantly improved method to saturate 't Hooft's upper limit.

\subsection{Ultralight axions}

\noindent
In the Introduction, we briefly discussed the QCD axion. In the canonical case both
the axion mass $m_a^{QCD}$ and its coupling to normal matter are
 inversely proportional to the the axion decay constant, $f_a$
 \begin{equation}
 m_a^{QCD} = 
 \frac{\sqrt{\chi_{QCD}}}{f_a}  \simeq m_{\pi}f_{\pi} \frac{\sqrt{m_u m_d}}{m_u + m_d} \left(\frac{1}{f_a}\right)
 \label{Maxion}
 \end{equation} 
where $\chi_{QCD}$ is the QCD topological susceptibility; $m_u,m_d$ and $m_{\pi}$ are the masses
of up-quark, down-quark, and $\pi$-meson respectively.

\bigskip

\noindent
The relations in Eq.(\ref{Maxion}) are inevitable when there is only one confining
QCD gauge group. The invisible axion mass is then within the range between 1 peV
and $1$meV. Rather, we are interested in an ULA which solves the strong CP problem
with suppressed $m_a$, if possible down to $\sim 10^{-33}$ eV, the minimum mass 
for which the axion Compton wavelength is smaller than the Hubble horizon.

\bigskip

\noindent
One way to accomplish this \cite{Hook1,Hook2,Ringwald1,Ringwald2} is to assume
that Nature is endowed with a discrete $Z_{{\cal N}}$ symmetry and ${\cal N}$
discrete worlds linked by the axion field. As we shall discuss, this leads to a suppression
of the canonical axion mass in Eq.(\ref{Maxion}) by a dependence for asymptotically large ${\cal N}$,
\begin{equation}
m_a \propto m_a^{QCD} \left(\frac{m_u}{m_d} \right)^{\frac{{\cal N}}{2}}
\label{suppression}
\end{equation}
Although Eq.(\ref{suppression}) is strictly valid only for large ${\cal N}$, it is a good
approximation already at low finite ${\cal N}$ which must be
an odd integer to solve the strong CP problem. To suppress from the visible axion
at $100$keV to the desired mass $10^{-33}$ eV, and given that $z\equiv m_u/m_d \simeq0.48$ we can estimate from Eq.(\ref{suppression}) that ${\cal N}=118.75 \rightarrow 119$, the nearest odd integer. Such an 
${\cal N} = O(100)$ theory
contains an $m_a \sim 10^{-33}$eV axion which solves the strong CP problem.

\bigskip

\noindent
Before studying the ULA contribution to the entropy of the universe, and to make our Letter
self-contained, we discuss a few details of the $Z_{{\cal N}}$ theory to explicate two of our
non-obvious assertions {\it ut supra}. It easiest to begin from ${\cal N}=2$ where there is only one degenerate
mirror world with the $Z_2$ symmetry between the two standard models SM and SM'
\begin{equation}
SM \leftrightarrow SM' ~~~~~ a\rightarrow a + \pi f_a
\label{Z2`}
\end{equation}
with the lagrangian
\begin{equation}
{\cal L} = {\cal L}_{SM} + {\cal L}_{SM'} + 
\frac{\alpha_s}{8\pi} \left( \frac{a}{f_a} - \theta \right) G_{\mu\nu}\tilde{G}^{\mu\nu} + 
\frac{\alpha_s}{8\pi} \left( \frac{a}{f_a} - \theta +\pi \right) G'_{\mu\nu}\tilde{G'}^{\mu\nu} + ...
\label{L2}
\end{equation}
where the ellipsis denotes $Z_{{\cal N}}$-symmetric portal couplings which may connect the
different mirror sectors. The axion potential is more shallow than for ${\cal N}=1$ because
of partial cancellation between terms for SM and SM', such as
\begin{equation}
\left[ \sqrt{ m_u^2 + m_d^2 + 2m_um_d {\rm cos}(\frac{a}{f_a})} \right.  \\
\left. + \sqrt{ m_u^2 + m_d^2 - 2m_um_d {\rm cos}(\frac{a}{f_a})} \right]
\label{V2}
\end{equation}

\noindent
In this case $\frac{a}{f_a}=0$ is an unstable maximum so strong CP is not solved. The origin
alternates between an unstable  maximum and a stable minimum for ${\cal N}=1,2,3,...$, thus explaining the
first of our two non-obvious assertions, that ${\cal N}$ must be an odd integer to solve the strong CP
problem.

\bigskip

\noindent
To understand the exponential suppression of the axion mass for large ${\cal N}$
of the form $m_a \sim z^{\frac{{\cal N}}{2}}$
requires a discussion of general ${\cal N}$.
With ${\cal N}$ copies of the SM the $Z_{{\cal N}}$ symmetry is 
\begin{equation}
SM_k \rightarrow SM_{k+1} ~~~~~ a \rightarrow \frac{2 \pi k}{{\cal N}} f_a ~~~~~~ 0 \leq k \leq ({\cal N} -1)
\label{ZN}
\end{equation}
and the lagrangian is
\begin{equation}
{\cal L}= \Sigma_{k=0}^{{\cal N}-1} 
\left[ {\cal L}_{SM_k} + \frac{\alpha_s}{8\pi}( \theta_a + \frac{2 \pi k}{{\cal N}} ) G_k \tilde{G}_k \right] + ...
\label{LN}
\end{equation}
The resulting axion potential is
\begin{equation}
V_{{\cal N}}(\theta_a) = -m_{\pi}^2f_{\pi}^2 ~ \Sigma_{k=0}^{{\cal N}-1} 
\sqrt{1 -\beta {\rm sin}^2 \left(\frac{\theta_a}{2}+\frac{\pi k}{{\cal N} }\right)}
\label{VNaxion}
\end{equation}
where $\beta = 4 m_u m_d (m_u+m_d)^{-2} = 4z(1+z)^{-2}$ with $z = m_u/m_d$.

\bigskip

\noindent
In the limit ${\cal N} \rightarrow \infty$, the leading term is
\begin{equation}
m_a^2 f_a^2 \sim 
\frac{m_{\pi}^2 f_{\pi}^2}{\sqrt{\pi}} \sqrt{\frac{1-z}{1+z}} {\cal N}^{\frac{3}{2}} z^{{\cal N}} \propto \left( \frac{m_u}{m_d} \right)^{{\cal N}}
\label{largeN}
\end{equation}
so that the ULA mass has the suppression $m_z \propto z^{\frac{{\cal N}}{2}}$ which was quoted without justification, {\it ut supra}. The asymptotic form in Eq.(\ref{largeN}) remains quite accurate even at the lowest
nontrivial odd integer ${\cal N}=3$. In the large ${\cal N}$ limit of Eq.(\ref{largeN}), the ULA becomes
massless as the Nambu-Goldstone boson of breaking a continuous
$U(1) \supset Z_{\infty}$.

\bigskip

\noindent
In the above, we showed how the QCD axion works for this specific ULA case. For the purpose of
model building, however, without introducing ${\cal N}$ copies, the same result can be obtained 
when we introduce a discrete symmetry
$Z_{{\cal N}}$ into the potential.

\bigskip

\noindent
Given an ultralight axion with mass $10^{-q}$ eV, solving the strong CP problem, we can estimate its
contribution to the entropy of the present universe by noting that the present mass of
the universe is $M_U \sim 10^{23} M_{\odot} \sim 10^{89} $eV. If the ULA comprises all the dark matter, it
contributes
\begin{equation}
S_U(t_0) \sim 10^{q-11} {\cal G}
\label{Saxion}
\end{equation}
so that for $q=33$ we come within one order of magnitude of saturating the 't Hooft holographic bound in Eq.(\ref{bounds}). In the previous subsection, we found this is not possible for PBHs. Based on our guiding principle,
we must conclude that ULAs are better suited to be the dark matters constituents than PBHs.

\bigskip

\noindent
From the point of view of entropy alone, any ultralight scalar or pseudoscalar can as well fulfil the
't Hooft holographic bound as can an ULA.

\section{Discussion of guiding principle}

\noindent
In Section 2, we enunciated our guiding principle as:"Subject to dynamical constraints, the dark matter constituents are such that
the entropy of the universe reaches its maximum possible value". This is remarkably similar to the
statement of the second law over 150 years ago by Clausius, the physicist who first coined the word {\it entropy}:
"The entropy of the universe tends to a maximum"(1865). It is interesting that he talked about the "universe" despite
the fact that his experience was limited to laboratory systems such as a box of ideal gas. Nevertheless,
by applying our guiding principle to dark matter in the real universe, we have succeeded to bring order out of the
chaos generated by very many candidates ranging in mass by over 117 OoMs, by reaching the more orderly selection that just two candidates, PBHs and ULAs, are consistent.

\bigskip

\noindent
A key ingredient in our analysis has been the insight by 't Hooft\cite{Hooft}
that the number of degrees of freedom (dofs) available for gravity in 3+1 spacetime dimensions
is not
greater than the number of dofs for quantum field theory in 2+1 spacetime dimensions.
His 1993 paper on gravity was seminal
to the holographic principle and to this Letter.

\bigskip

\noindent
For PBHs of mass $10^p M_{\odot}$ and ULAs of mass
$10^{-q} eV$ respectively, taken as constituting all dark matter, our results
for the present entropy of the universe are
\begin{equation}
S_U(t_0) \sim 10^{1+p} {\cal G} ~~~ {\rm and} ~~~ S_U(t_0) \sim 10^{q-11} {\cal G}.
\label{PBHsUAs}
\end{equation}
In the former, even such a gigantic SLAB as $p=18$ falls a few OoMs short;
in the latter, an ultralight axion at the Hubble scale $q=33$ can fully saturate the
holographic bound, and it is therefore better suited as the constituent of dark matter.

\bigskip

\noindent
That being said, $DM=PIMBHS$ \cite{PaulFrampton} remains a
real possibility and merits further study using
microlensing of stars in the Magellanic Clouds\cite{MACHO}. 
Such an experiment employing DECam data was recently attempted at LLNL but
stymied \cite{DECam} by crowded field issues. A further opportunity to 
microlens may occur at the Rubin Observatory\cite{VeraCRubinObservatory}.

\bigskip

\section*{Acknowlegements}
CC is partially supported by INFN of Italy under the QFT-HEP initiative. JEK is supported in part by the National Research Foundation grant NRF-2018R1A2A3074631.

\end{document}